\documentclass[twocolumn, showpacs, longbibliography, prx, aps, amsmath, amssymb,superscriptaddress]{revtex4-1}
\usepackage{amsmath} % needed for \tfrac, \bmatrix, etc.
\usepackage{amsfonts} % needed for bold Greek, Fraktur, and blackboard
\usepackage{graphicx} % needed for figures
\usepackage[colorlinks,linkcolor=blue,citecolor=blue,urlcolor=blue]{hyperref}
\usepackage{float}
\usepackage[normalem]{ulem}
\usepackage[switch,columnwise]{lineno}

\begin{document}

\title{\Large{Non-reciprocity and zero reflection in nonlinear cavities with tailored loss }}

\author{S.R.K. Rodriguez} \email{s.rodriguez@amolf.nl}

\affiliation {Center for Nanophotonics, AMOLF, Science Park 104, 1098 XG Amsterdam, The Netherlands}

\author   {V. Goblot}
\affiliation {Centre de Nanosciences et de Nanotechnologies, CNRS, Universit\'{e} Paris-Sud, Université Paris-Saclay, 91120 Palaiseau, France}

\author   {N. Carlon Zambon}
\affiliation {Centre de Nanosciences et de Nanotechnologies, CNRS, Universit\'{e} Paris-Sud, Université Paris-Saclay, 91120 Palaiseau, France}

\author   {A. Amo}
\affiliation {Univ. Lille, CNRS, UMR 8523 -- PhLAM -- Physique des Lasers Atomes et Mol\'{e}cules, F-59000 Lille, France}

\author   {J. Bloch}
\affiliation {Centre de Nanosciences et de Nanotechnologies, CNRS, Universit\'{e} Paris-Sud, Université Paris-Saclay, 91120 Palaiseau, France}

\date{\today}

\begin{abstract}

We demonstrate how to tailor the losses of nonlinear cavities in order to suppress their reflection and enhance their non-reciprocal transmission. We derive analytical expressions predicting the existence of zero-reflection channels in single and coupled nonlinear cavities, depending on the driving frequency and loss rates. While suppressing the reflection from a single cavity imposes a stringent condition on the input-output leakage rates, we demonstrate that this condition can be significantly relaxed in systems of coupled cavities. In particular, zero-reflection and non-reciprocity can be achieved across a range of driving frequencies in coupled cavities by tuning the output leakage rate alone. Numerical calculations based on the driven-dissipative Gross-Pitaevksii equation, usually employed to describe microcavity polaritons, reveal the spatial phenomenology associated with zero-reflection states and provide design guidelines for the construction of nonlinear optical isolators.

\end{abstract}

\maketitle

\begin{figure}[hbtp]
 \centerline{{\includegraphics[width=1\linewidth]{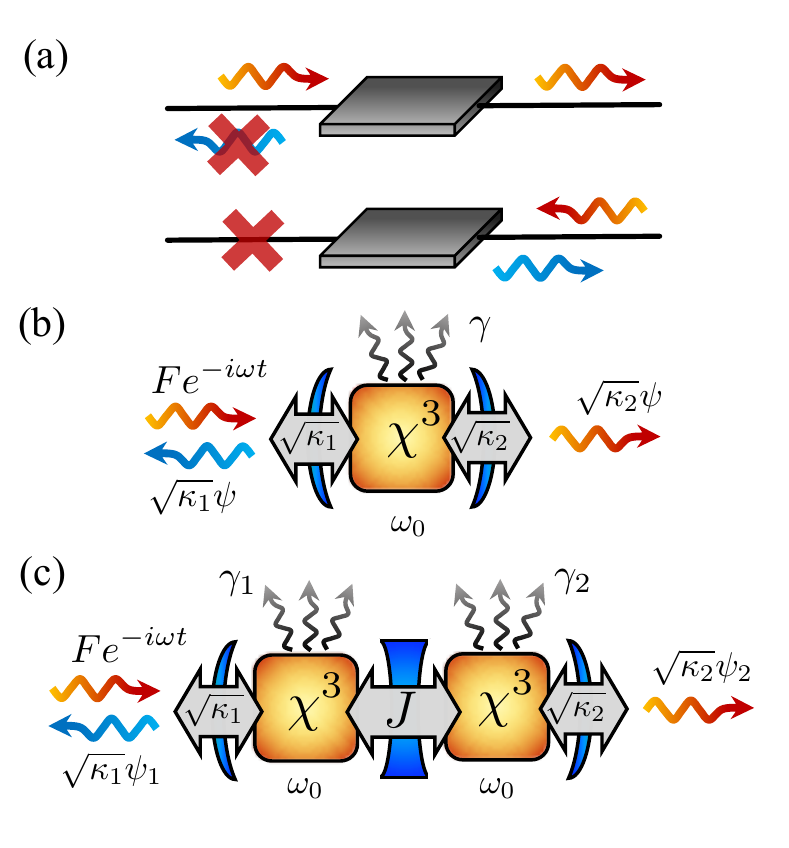}}}\caption{(a) Ideal features of an optical isolator: zero reflection and high transmission for forward propagation (top panel), and zero transmission for backward propagation (bottom panel). (b) Sketch of a single mode cavity with Kerr-type nonlinearity, intrinsic loss $\gamma$, and coupled to two separate input-output ports at rate $\kappa_{1,2}$. (c) Sketch of two coupled cavities [each one as in (b)] with coupling energy $J$. In both (b) and (c), the resonant system is driven from the left (port 1) by a monochromatic field of frequency $\omega$ and amplitude $F$. For testing the reciprocity of the system, the driving field is switched to the right (not shown).}
\label{fig1}
\end{figure}

Lorentz reciprocity, which in the absence of gain and loss is equivalent to time reversal symmetry~\cite{Carminati2000}, states that the relationship between source and detector remains unchanged when their positions are exchanged. Reciprocity holds for linear time-invariant systems with symmetric permittivity and permeability tensors~\cite{Jalas13}. Systems not constrained by Lorentz reciprocity are of interest in many fields, particularly in photonics since they may function as optical diodes or isolators~\cite{Ramezani10, Fan12Science, Lipson12, Fan13OL, Baets13, Sounas13, Tzuang14, Chang14, Peng14, Alu14, Mork15, Clerk15, Shen16, Ruesnik16, Sounas17, Lawrence18, Sounas18}. Figure~\ref{fig1} illustrates two main features of an ideal optical isolator: light propagates one way only, and reflection at the input port is zero.

Recently, many efforts have concentrated on the design of compact magnetic-free non-reciprocal systems. One approach to non-reciprocity is based on dynamic modulation --- a departure from the time-invariance assumption on which Lorentz reciprocity relies~\cite{Yu09, Lipson12, Fang12, Fang12NP, Sounas13, Alu14, Sounas17NP}. An alternative approach to non-reciprocity is based on the combination of nonlinearity and spatial symmetry breaking~\cite{Ramezani10, Chang14, Peng14, Sounas17, Sounas18}. While every approach to non-reciprocity offers benefits and limitations~\cite{Shi15, Mann18}, a common drawback of many approaches is that reflection from the input port tends to be deleteriously high at non-reciprocal conditions; see for example Table 1 of Ref.~\cite{Fan13OL}, comparing insertion losses and non-reciprocity for several systems.

Here we take a new approach to simultaneously achieve non-reciprocity and zero reflection from the input port of dissipative cavities with Kerr-type nonlinearity. Our approach relies on tailoring the leakage rates of the cavities to their input-output ports. In contrast to previous approaches combining Fano and Lorentzian resonances with a suitable delay line in-between~\cite{Sounas18}, our method works for coupled cavities even when their eigenfrequencies and intrinsic loss rates are equal. To benchmark our results, we first provide a detailed analysis of non-reciprocity in a single nonlinear cavity with separate input-output ports.  In the presence of intrinsic cavity losses, non-reciprocity with zero reflection can only be achieved for a particular value of the input-output leakage rate difference. As we will show, this stringent condition can be relaxed in systems of coupled cavities. We find analytical expressions for  the conditions giving zero reflection in  non-reciprocal systems of single and coupled cavities. In addition, through numerical calculations based on the driven-dissipative Gross-Pitaevskii equation, we present a design for a realistic semiconductor polariton system where our predictions can be experimentally tested.

\section{Single nonlinear cavity}\label{sec:SingleKR}

Figure~\ref{fig1}(b) illustrates the system studied in this section: a single mode cavity with resonance frequency $\omega_0$, intrinsic loss rate $\gamma$, and a $\chi^3$ Kerr-type nonlinearity leading to photon-photon interactions of strength $U$. The cavity is coupled to two separate input-output ports at rates $\kappa_{1,2}$. A monochromatic field of frequency $\omega$ and amplitude $F$  drives the cavity through port 1. Within the mean-field approximation neglecting quantum fluctuations~\cite{Drummond80, WallsMilburn}, the cavity field amplitude $\psi$ obeys the following equation of motion ($\hbar=1$):
\begin{equation}\label{eq:NonlinearShrodEq}
i\dot{\psi}=\left( \omega_0 - i \frac{\Gamma}{2} + U \vert \psi \vert^2 \right)\psi + i \sqrt{\kappa_{1}}F	e^{-i\omega t},
\end{equation}
\noindent where $\Gamma=\gamma+\kappa_1+\kappa_2$ is the total loss. The steady-state solutions are found by setting $\dot{\psi}=0$, and inserting the ansatz $\psi(t)=\psi e^{-i\omega t}$ in Eq.~\eqref{eq:NonlinearShrodEq}. For convenience, we move to a frame rotating at the driving frequency $\omega$. In this rotating frame, the detuning $\Delta=\omega-\omega_0$ is the relevant energy parameter and $F	e^{-i\omega t} \rightarrow F$. The steady-state field is then a solution to the following algebraic equation:
\begin{equation}\label{eq:SSequationSKR}
0=\left( -\Delta - i \frac{\Gamma}{2} + U \vert \psi \vert^2 \right)\psi + i \sqrt{\kappa_{1}}F.
\end{equation}
Once $\psi$ is obtained, the transmittance $\mathcal{T}$ and reflectance $\mathcal{R}$ can be calculated as follows:
\begin{align}
	\mathcal{T}_1 & = \left \vert \frac{\sqrt{\kappa_2} \psi}{F} \right \vert^2= \frac{\kappa_1\kappa_2}{(\Delta-U n)^2+\Gamma^2/4}, \label{T1cav} \\
	\mathcal{R}_1 & = \left \vert \frac{F-\sqrt{\kappa_1} \psi}{F} \right \vert^2= 1-\frac{\kappa_1(\kappa_2+\gamma)}{(\Delta-U n)^2+\Gamma^2/4}, \label{R1cav}
	\end{align}
with $n=\vert\psi\vert^2$ the number of photons in the cavity. The subscript $^\backprime 1 ^\prime$ of $\mathcal{R}$ and $\mathcal{T}$ indicates that the cavity is driven through port 1. Similar expressions can be obtained for $\mathcal{R}_2$ and $\mathcal{T}_2$ by letting $1 \rightarrow 2$ and $2 \rightarrow 1$, and solving Eq.~\eqref{eq:SSequationSKR} again for $n$. Notice that if $\gamma\neq 0$, $\mathcal{R}+\mathcal{T}<1$

While equations \eqref{T1cav} and \eqref{R1cav} seem to be symmetric in $\kappa_{1}$ and $\kappa_{2}$, the response of the system is not the same when driven through port 1 and port 2 because of the nonlinear term $U n$ . Indeed, for $\kappa_1 \neq \kappa_2$, driving through the $i^{th}$ port instead of through the $j^{th}$ ports rescales the effective drive amplitude  by $\sqrt{\kappa_i}/\sqrt{\kappa_j}$. For a fixed $F$, this results in a different $n$ when driving through port 1 and 2. Consequently, the cavity transmission is nonreciprocal when the input ports are switched.

Notice that whereas the transmittance is simply proportional to $|\psi|^2$, the reflectance is determined by the interference between the driving field and the cavity field. In particular, $\mathcal{R}_1$ vanishes when $F=\sqrt{\kappa_1}\psi$, corresponding to total destructive interference between the two fields. Substituting this relation in Eq.~\eqref{eq:SSequationSKR}, we find two necessary conditions for $\mathcal{R}_1=0$:
\begin{equation}\label{eq:ZeroReflection_SKR}
	\begin{aligned}
	\kappa_{1} &= \kappa_2 + \gamma = \Gamma/2 ,\\
	F^2_c & =\kappa_{1} \Delta / U,
	\end{aligned}
\end{equation}

\noindent with $F_c$ the critical driving amplitude for which $\mathcal{R}_{1,c} = 0$. Hereafter, the quantities evaluated at $F=F_c$ will have the `c' subscript. If the conditions in Eq. \eqref{eq:ZeroReflection_SKR} hold, the transmittance is $\mathcal{T}_{1,c}=\kappa_{2}/\kappa_{1}$. This leads to the following conclusion: unitary transmission requires $\gamma=0$ and $\kappa_1=\kappa_2$. The latter would imply that the device is mirror-symmetric and, therefore, it would not show any non-reciprocity. Hence, in this configuration, either unitary transmission or non-reciprocity can be achieved, but not both simultaneously. \\

\begin{figure}[t]
 \centerline{{\includegraphics[width=\linewidth]{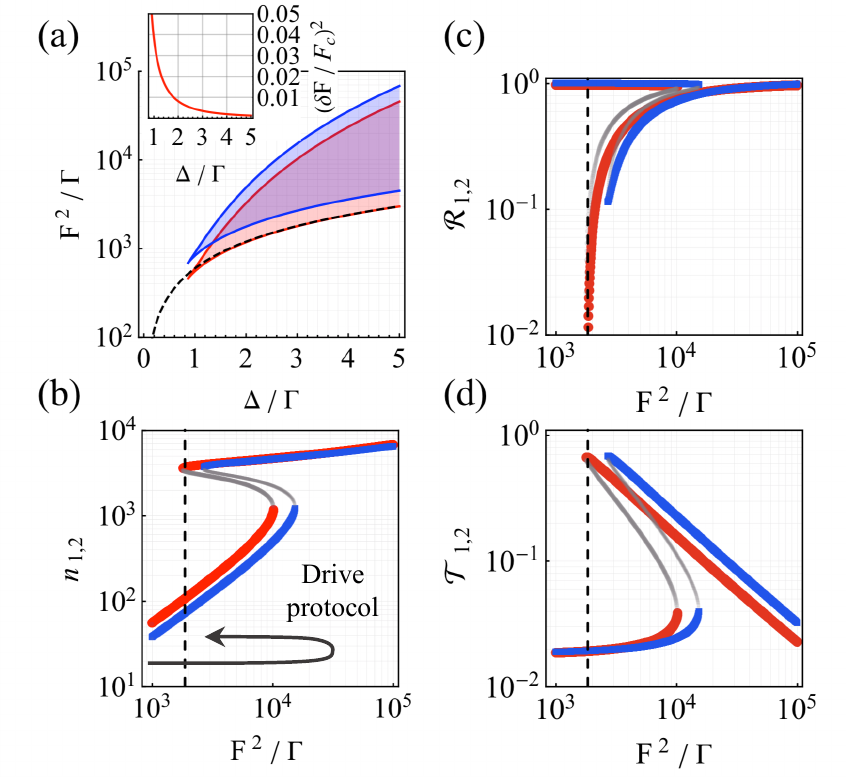}}}\caption{Calculations for a single cavity as depicted in Fig.~\ref{fig1}(b), with $U=0.005\gamma$, $\kappa_1=\Gamma/2$, $\kappa_2=\Gamma/3$, and $\gamma=\Gamma/6$. (a) Red (blue) curves enclose the range of normalized driving power $F^2/\Gamma$ and normalized laser-cavity detunings $\Delta/\Gamma$, with  $\Delta=\omega-\omega_0$, giving rise to bistability when driving through port 1 (port 2). The inset shows the largest power fluctuation $(\delta F / F_c)^2$ the system can withstand without losing its non-reciprocity by falling from the bistability branch with largest number of photons; $F_c$ is the critical driving amplitude at which  $\mathcal{R}_{1}=0$.  (b), (c), and (d) Number of photons in the cavity, reflectance, and transmittance, respectively, all for $\Delta/\Gamma=3$. The vertical black line in (c) and (d) indicates the critical drive power for which the reflectance vanishes, as predicted by Eq. \eqref{eq:ZeroReflection_SKR}.}
\label{fig2}
\end{figure}

To quantify the non-reciprocity, we need to define an appropriate figure of merit. Non-reciprocity has been previously assessed through the ratio of forward-to-backward transmission at a fixed intensity~\cite{Fan12Science, Fan13OL}. Alternatively, an isolation intensity range can be defined as the ratio of input intensities from opposite propagation directions that lead to the same transmission~\cite{Sounas18}. Applying these definitions to nonlinear systems exhibiting bistability or multistability is non-trivial. A bistable cavity sustains two stable steady-states with different photon numbers at the same driving conditions~\cite{Gibbs}. The observed steady-state depends on the driving history of the system; Hysteresis emerges as a driving parameter is scanned across a bistability~\cite{Gibbs76, Rodriguez17}. Since $\mathcal{T}$ cannot be uniquely defined within the hysteresis range, the isolation ratio is also not uniquely defined. In principle, this ambiguity can lead to asymmetric forward-to-backward transmission at fixed driving conditions if the system is biased into different states, even for symmetric systems. To avoid this ambiguity, we propose to evaluate $\mathcal{T}$ not only at equal $F$, but also at equal driving histories. For instance, if the cavity is driven across a hysteresis cycle in the forward direction, then the same driving protocol should be followed in the backward driving direction. For bistable cavities, this criterion gives the `worst possible' isolation ratio $IR$ as follows:
\begin{equation}\label{eq:IsolationRatio}
IR= \frac{\mathcal{T}_{1,max}}{ \mathcal{T}_{2,max}}.
\end{equation}
\noindent $\mathcal{T}_{1,max}$ and $\mathcal{T}_{2,max}$ are the maximum transmittance one can obtain when driving the system from port $1$ or $2$, respectively. For $\kappa_1 = \kappa_2$, one finds using Eq.~\eqref{T1cav} that $IR=1$, i.e., the system is reciprocal. Meanwhile, for asymmetric systems ($\kappa_1 \neq \kappa_2$) with more than one possible forward-to-backward transmission ratio, our definition of $IR$ gives the value closest to one, i.e., the worst possible isolation ratio for a given input power.\\

Let us now analyze the non-reciprocal behavior of a single bistable cavity with $\kappa_1 \neq \kappa_2$. For repulsive interactions $U>0$, bistability occurs for $\Delta >\sqrt{3}\Gamma / 2$. Within the hysteresis cycle range, the intracavity photon number $n$ is constrained by $n_{-}\leq n \leq n_{+}$, with $n_{\pm}=(2\Delta)/(3 U)\pm(6U)^{-1}\sqrt{4\Delta^2-3\Gamma^2}$. The red (blue) shaded areas in Fig. \ref{fig2}(a) enclose the values of $F^2/\Gamma$ and $\Delta/\Gamma$ where bistability takes place when driving through port 1 (port 2). Here and throughout the manuscript, the same color code will be used for quantities computed when driving through the two ports. The shape of the bistability region is the same when driving through ports 1 and 2, but this region is shifted to higher $F$ when driving through port 2 because $\kappa_2 < \kappa_1$.

In Fig.~\ref{fig2}(b) we plot $n$ versus $F^2/\Gamma$ for a weakly nonlinear cavity $U\ll\gamma$ at the detuning $\Delta=3\Gamma$, where the system displays a bistable behavior. The stability of the solutions was assessed by evaluating the spectrum of small fluctuations around the steady-state~\cite{Drummond80}. Unstable solutions are marked with gray dots. As we have set $\kappa_1=\Gamma/2 = \kappa_2 + \gamma$ to satisfy Eq.~\eqref{eq:ZeroReflection_SKR} for the calculations in Fig. \ref{fig2}(b), at fixed $F$ and for finite $\gamma$, the effective driving strength through port 1 is greater than through port 2. Consequently, the bistability range is shifted in $F$ and the transmission is non-reciprocal.

Figures~\ref{fig2}(c) and~\ref{fig2}(d) show $\mathcal{R}_{1,2}$ and $\mathcal{T}_{1,2}$, respectively, corresponding to the steady-state solutions in Fig. \ref{fig2}(b). Notice in Fig.~\ref{fig2}(c) the sharp dip in $\mathcal{R}_1$. To access this state, one needs to apply the drive protocol sketched in Fig. \ref{fig2}(b), to reach the high $n$ state very close to the bistability falling edge (not jumping down) when driving through port 1. $\mathcal{R}_1=0$ at the power indicated by the dashed line in Fig.~\ref{fig2}(c), which is the critical drive amplitude $F_c =\kappa_{1} \Delta / U$ predicted by Eq.~\eqref{eq:ZeroReflection_SKR}. A finite $\mathcal{R}_1$ is observed in the numerical calculations because of the finite step size in $F$. Figure~\ref{fig2}(d) shows that for $\mathcal{R}_1=0$, $\mathcal{T}_1=1/2$, and $\mathcal{T}_2\ll 1$: driving through port 2 at the same $F$ can only set the system in the low-$n$ branch, such that IR$\approx 36$. In this regime, starting from Eq.~\eqref{T1cav}, we can deduce an approximate formula for the isolation ratio at $F_c$,
\begin{equation}\label{eq:IsolationRatio_SKR}
	IR_c\approx 1+ \frac{4\Delta^2}{\Gamma^2},
\end{equation}
 where we used the fact that $\mathcal{T}_{2,c}\approx \kappa_1\kappa_2/(\Delta^2+\Gamma^2/4)$ because the term $U n$ is negligible in the low $n$ steady-state. Interestingly, in the regime we are considering in which the driving field intensity is in between the two down-falling bistability edges shown in Fig.~\ref{fig2}(b), the $IR$ does not depend on the relative values of $\kappa_{1,2}$ and $\gamma$; it only depends on the ratio of the detuning $\Delta$ to the total loss $\Gamma$.\\

\begin{figure*}[t]
 \centerline{{\includegraphics[width=\linewidth]{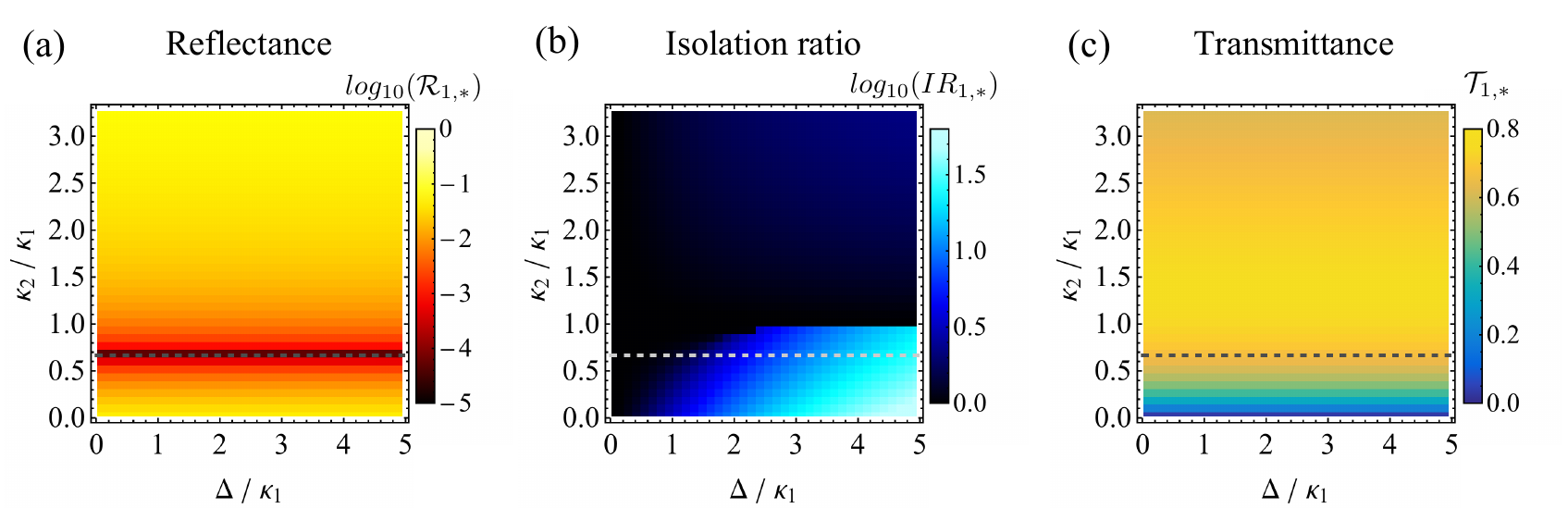}}}\caption{ Calculations for a single cavity as depicted in Fig.\ref{fig1}(b), with $U=0.005\gamma$, $\gamma=\kappa_1/3$ and $\kappa_2 / \kappa_1= 2/3$. (a), (b), (c): reflectance, isolation ratio, and transmittance in color scale as a function of the leakage rate ratio $\kappa_2/\kappa_1$ and normalized laser-cavity frequency detuning $\Delta/\kappa_1$. The horizontal dashed line indicates the value of $\kappa_2/\kappa_1$ leading to zero-reflection, as predicted by Eq.~\eqref{eq:ZeroReflection_SKR}.}
\label{fig3}
\end{figure*}

Vanishing $\mathcal{R}_1$ and nearly unitary transmission can be achieved in the limit of arbitrarily small but finite $\gamma$ as long as $\kappa_{1}= \kappa_2 + \gamma$. However, this makes it increasingly difficult to unidirectionally bias the system into the desired bistable state because the difference between the left- and right-driven bistability threshold diminishes when $\gamma$ vanishes. A similar argument holds for the isolation ratio. Even if $IR_c$ can be made arbitrarily high by increasing $\Delta/\Gamma$, the difference $\delta F$ between $F_c$ and the value of $F$ corresponding to the falling-edge of the high $n$ steady-state becomes increasingly small. Consequently, tiny fluctuations in the input power ($\propto F^2$) have an increasing probability of making the cavity switch to the low $n$ steady-state where there is poor isolation and high reflectance. A relevant figure of merit for practical implementations is the ratio $(\delta F/ F_c)^2$. This ratio quantifies the largest power fluctuation which the system can withstand without losing its non-reciprocity. We plot this quantity in the inset of Fig.~\ref{fig2}(a) as a function of $\Delta/\Gamma$. For $\Delta/\Gamma=3$ as considered in Figs.\ref{fig2}(b,c,d), power fluctuations of $0.5\%$ will spoil both the isolation and low-reflectance of the system.\\

We now seek an expression for the maximum transmission $\mathcal{T}_{1,*}$ (or correspondingly, minimum reflection $\mathcal{R}_{1,*}$) which can be achieved at a certain $\kappa_{1,2}$ and $\gamma$, while driving through port 1.
From Eqns.~\eqref{T1cav},\eqref{R1cav}, we see that this is achieved for $\Delta = Un$. The drive amplitude $F_*$ yielding $\mathcal{T}_{1,*}, \mathcal{R}_{1,*}$ can also be determined by plugging the latter condition in equation \eqref{eq:SSequationSKR}. We get:
\begin{equation}\label{eq:Best_RTF}
	\begin{aligned}
	\mathcal{T}_{1,*} & =\frac{4\kappa_{1}\kappa_{2}}{\Gamma^2}\\
	\mathcal{R}_{1,*} & = \frac{(\gamma-\kappa_{1}+\kappa_{2})^2}{\Gamma^2}\\
	F^2 _*&=\frac{\Gamma^2}{4}\frac{\Delta}{\kappa_{1}U}.
	\end{aligned}
\end{equation}

The isolation ratio can be obtained by computing $\mathcal{T}_2$ at the drive amplitude $F_*$. These results, together with Eq.~\eqref{eq:IsolationRatio_SKR}, completely determine the best achievable performance in terms of $\mathcal{R}$, $\mathcal{T}$, and $IR$, as well as the driving amplitude at which this condition manifests, for any set of parameters $(\Delta,U,\kappa_{1,2},\gamma)$.\\

To illustrate how the performance degrades when  departing from the condition $\kappa_{1}= \kappa_2 + \gamma$, we compute and report in Fig.~\ref{fig3} $\mathcal{R}_{1,*}$, $\mathcal{T}_{1,*}$, and $IR$ as a function of $\kappa_2$ and $\Delta$ for $\gamma=0.02$ meV and $\kappa_1 = 0.06$ meV. These are experimentally relevant parameters as explained ahead. The horizontal dashed line in all panels indicates the value of $\kappa_2$ for which $\mathcal{R}_1=0$. Increasing $\kappa_2$ above this ideal value degrades the system in terms of $\mathcal{R}_{1,*}$, and $IR$. Decreasing $\kappa_2$ below the ideal value improves $IR$, but degrades the performance both in terms of $\mathcal{R}_{1,*}$ and $\mathcal{T}_{1,*}$. In contrast, as we will see in section II, tuning $\kappa_2$ in coupled cavities allows $\mathcal{R}_1=0$, high $\mathcal{T}_1$, and high $IR$, at variable values of $\Delta$.\\

In summary, for single Kerr resonator, either unitary transmission ($\gamma=0$) or non-reciprocity can be achieved. If we allow finite losses ($\gamma\neq0$) and the mirror symmetry of the system is broken ($\kappa_{1}\neq\kappa_{2}$), a nonlinear steady state with $\mathcal{R}_1=0$ and an isolation ratio growing quadratically with $\Delta/\Gamma$ (see Eq.~\eqref{eq:IsolationRatio_SKR}) can be achieved. However to achieve this effect, strict conditions have to be met for the ratios of the input-output couplings $\kappa_{1,2}$ with respect to the losses $\gamma$, which may limit the performance of realistic implementations of the scheme. Such strict conditions can be relaxed in systems of two coupled resonators.

\section{Two coupled nonlinear cavities}

In this section we consider two mutually coupled cavities as depicted in Figure~\ref{fig1}(c). Notice that ports 1 and 2 are now connected to different cavities. As we will show, this allows relaxing the stringent conditions on the loss rates leading to zero-reflection at the input port while still maintaining high non-reciprocity.

In a frame rotating at the driving frequency $\omega$, the equations for the coupled cavity fields are:
\begin{align}
\left( -\Delta_1 - i \frac{\Gamma_1}{2} + U n_1 \right)\psi_1 - J \psi_{2} &=-i\sqrt{\kappa_{1}}F \label{cav1}\\
\left( -\Delta_2 - i \frac{\Gamma_2}{2} + U n_2 \right)\psi_2 - J \psi_{1} &= 0           \label{cav2}
\end{align}
\noindent with $\psi_j$ the field, $\Delta_{j}=\omega-\omega_j$ the laser-cavity detuning, $\Gamma_{j} = \gamma + \kappa_{j}$ the total loss, and $n_j = |\psi_j|^2$ the number of photons, in the $j^{th}$ cavity ($j=1,2$). $J$ is the coupling between the 2 cavities. Calculating $\psi_j$ (see Appendix B for details) allows us to get the steady-state photon numbers $n_j$ and to assess the stability of the steady-states~\cite{Sarchi08}. For brevity, we omit details of the stability analysis which can be found in Ref.~\cite{Sarchi08}. We only recall that the coupled equations~\eqref{cav1} and~~\eqref{cav2} admit multiple steady-states at certain driving conditions, i.e., multi-stability. Each state can be classified as: (i) stable, (ii) single-mode unstable, or (iii) parametrically unstable~\cite{Sarchi08}. Here, we are interested in finding stable steady-states leading to minimum reflection at the input port and high transmission.

The transmittance $\mathcal{T}$ and reflectance $\mathcal{R}$, when driving through port 1, can be defined as follows:
\begin{align}
	\mathcal{T}_1 & = \left \vert \frac{\sqrt{\kappa_2} \psi_2}{F} \right \vert^2 \label{eq:T2cavs}\\
	\mathcal{R}_1 & = \left \vert \frac{F-\sqrt{\kappa_1} \psi_1}{F} \right \vert^2 \label{eq:R2cavs}
\end{align}
Comparing Eqns.~\eqref{eq:T2cavs} and~\eqref{eq:R2cavs} with Eqns.~\eqref{T1cav} and~\eqref{R1cav} reveals an important feature of coupled cavities with respect to a single cavity. The cavity field responsible for $\mathcal{R}_1$, namely $\psi_1$, is not the field responsible for maximizing $\mathcal{T}_1$, namely $\psi_2$. This feature opens new possibilities to achieve $\mathcal{R}_1=0$ and high non-reciprocity over extended parameter ranges.

Next, we pose the following question: Given a pair of identical cavities with eigenfrequency $\omega_0$, intrinsic loss $\gamma$, and mutual coupling $J$, for which values of the parameters $\Delta=\Delta_1=\Delta_2$ and $\kappa_{1,2}$ can we observe $\mathcal{R}_1=0?$ Our question is relevant to optical experiments, where $\Delta$ and $\kappa_{1,2}$ are typically external parameters which can be adjusted in-situ. $\Delta$ can be adjusted with a tunable laser, while $\kappa_{1,2}$ can be adjusted in evanescently coupled cavity-waveguide systems by tuning the cavity-waveguide distance, for example.

In Appendix C we address the above question by deriving analytical expressions guaranteeing the existence of a zero-reflection state. Unlike for a single cavity, we find two solutions giving $\mathcal{R}_1=0$ for coupled cavities. One of these solutions is equivalent to Eq.~\eqref{eq:ZeroReflection_SKR} (details ahead). The additional solution guaranteeing $\mathcal{R}_1=0$  in the coupled cavity case reads:

\begin{equation} \label{eq11V}
\Delta^2 = (\kappa_1 + \kappa_2)^2 \left( \frac{J^2}{(\kappa_1 - \gamma)(\kappa_2 + \gamma) } - \frac{1}{4} \right).
\end{equation}

\noindent In the derivation of Eq.~\eqref{eq11V}, we also find that $\mathcal{R}_1=0$ requires $n_2 \Gamma_2 = n_1 \tilde{\Gamma_1}$, with $\tilde{\Gamma_1} = \kappa_1 - \gamma$. This result demonstrates the key role that the losses play in achieving $\mathcal{R}_1=0$ by controlling the power flow through the system and fixing the relative number of photons in the two cavities.

\begin{figure}[!t]
 \centerline{{\includegraphics[width=0.9 \linewidth]{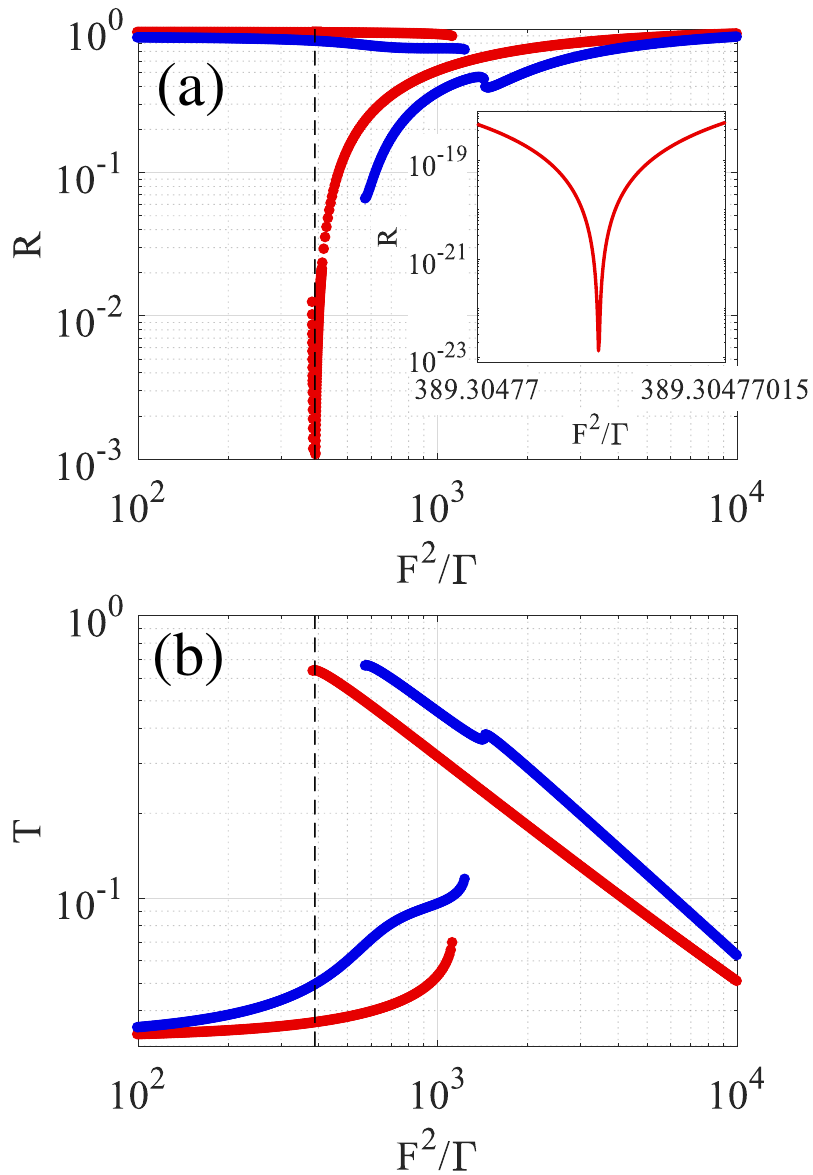}}}\caption{ Reflectance (a) and Transmittance (b) calculated for two coupled cavities as depicted in Fig.~\ref{fig1}(c), with $J=0.1$ meV, $\gamma=0.02$ meV, $U=0.07$ $\mu$eV, $\kappa_1 = 0.06$ meV, and $\kappa_2 = 0.4494$ meV. The dashed line in both panels indicates the critical driving power for which the reflectance vanishes. The inset of (a) is a zoom into the reflectance dip.}
\label{fig4}
\end{figure}

Next, we numerically solve equations~\eqref{cav1} and~\eqref{cav2} and calculate $\mathcal{R}_1$ and $\mathcal{T}_1$ as a function of $F^2/\Gamma$. Figures~\ref{fig4}(a,b) show $\mathcal{R}_1$ and $\mathcal{T}_1$ for values of $J$, $U$, $\gamma$ and $\kappa_{1,2}$, given in the caption. These values correspond to an experimentally realizable configuration to be discussed in the next section. The detuning was set to $\Delta=4.513 \kappa_1$ to satisfy Eq.~\eqref{eq11V} in combination with the other parameters.

Figure~\ref{fig4}(a) displays a sharp dip in $\mathcal{R}_1$ at $F^2/\Gamma=389.3$. The inset of Fig.~\ref{fig4}(a) shows a zoom into the dip, evidencing that $\mathcal{R}_1$ is suppressed by $\sim23$ orders of magnitude, limited by machine precision. At the driving power for which $\mathcal{R}_1=0$, non-reciprocity with $IR=13$ is obtained [see Fig.~\ref{fig4}(b)] .  The small jump in $\mathcal{R}_2$ and $\mathcal{T}_2$ around $F^2/\Gamma = 1450$ in the main panel is associated with an additional bistability. Cascades of bistabilities and multistabilities emerging when driving one of two coupled cavities have been previously studied~\cite{Sarchi08}, and experimentally observed~\cite{Deveaud16, Rodriguez16}.

\begin{figure*}[!t]
 \centerline{{\includegraphics[width=\linewidth]{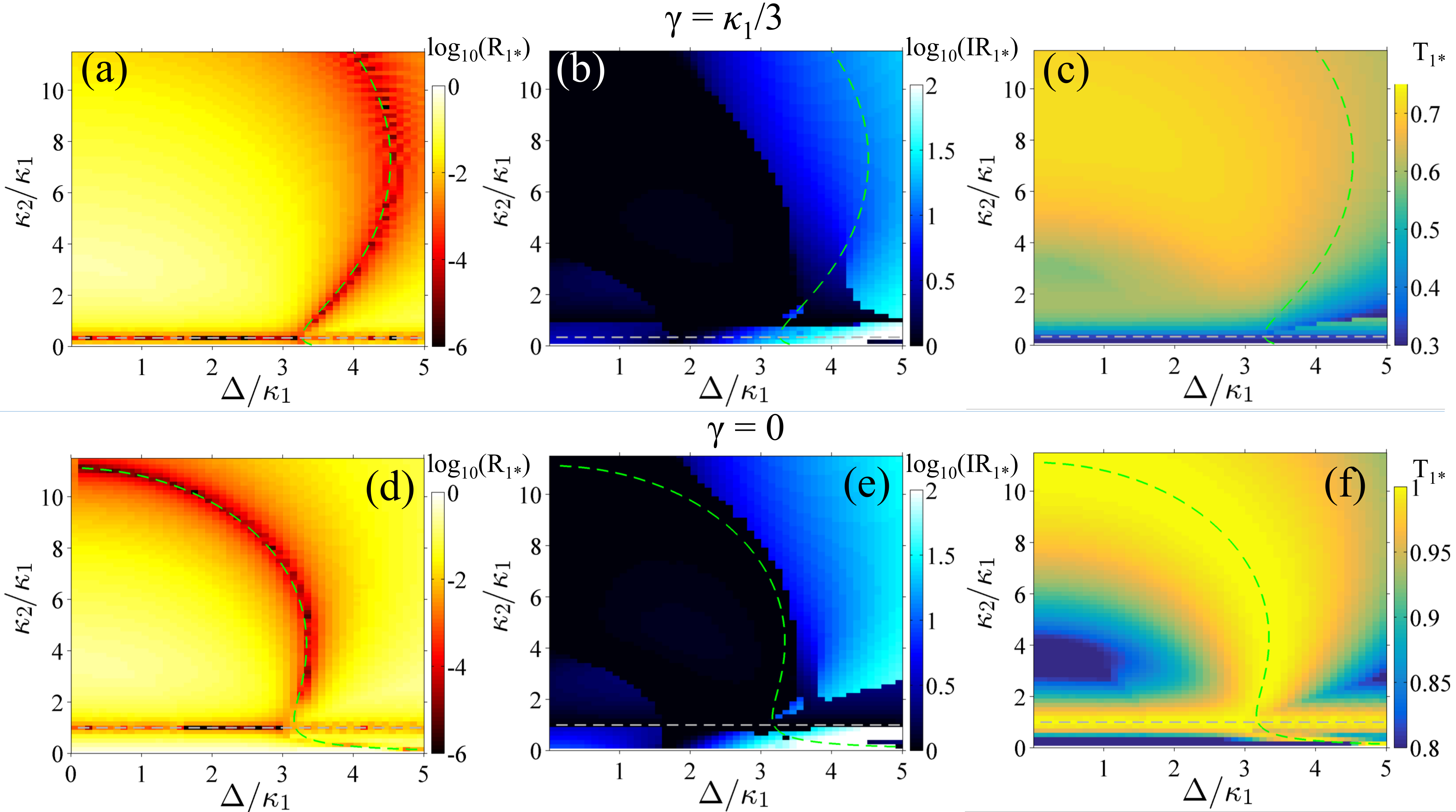}}}\caption{ Calculations for two coupled cavities as depicted in Fig.~\ref{fig1}(c), with $J=0.1$ meV, $U=0.07$ $\mu$eV, and $\kappa_1 = 0.06$ meV. In panels (a,b,c) $\gamma = \kappa_1/3$, and in panels (d,e,f) $\gamma =0$. Panels (a,d) show the minimum reflectance $\mathcal{R}_{1,*}$ observed at any driving power $F^2$. Panels (b,e) show the isolation ratio $IR$ at the same power for which $\mathcal{R}_{1,*}$ was observed. Panels (c,f) show the transmittance $\mathcal{T}_{1,*}$ of the same state associated with $\mathcal{R}_{1,*}$. Note that $\mathcal{R}_{1,*}$ and $IR$ are plotted in log scale, while $\mathcal{T}_{1,*}$ is plotted in linear scale. The green dashed curves in all panels are analytical predictions for zero-reflection from Eq.~\eqref{eq11V}; this state is exclusive to coupled cavities. The gray dashed curves in all panels are analytical predictions for zero-reflection based on Eq.~\eqref{eq:ZeroReflection_SKR}, but letting $\gamma \rightarrow 2\gamma$ because we have two dissipative cavities; this state corresponds to effective single-cavity behavior. }
\label{fig5}
\end{figure*}

Next, we assess $\mathcal{R}$, $\mathcal{T}$, and the isolation ratio $IR$ [Eq.~\eqref{eq:IsolationRatio}], for systematic variations of the coupled cavity system parameters. To this end, we first calculate $\mathcal{R}$ and $\mathcal{T}$ over a wide range of $F$ for a system with fixed $\Delta$ ,$U$, $\kappa_{1,2}$, $\gamma$, and $J$. We perform this calculation first driving through port 1, and then driving through port 2. The $F$-scan starts at low values for which the system is in the linear regime, and ends at high values which are well above all nonlinear thresholds. We then search for $\mathcal{R}_{1,*}$, i.e. the minimum value of $\mathcal{R}_1$, and estimate the corresponding value  of the transmission $\mathcal{T}_{1,*}$.  We also evaluate $IR$ at the power $F_*$ corresponding to $\mathcal{R}_{1,*}$. The results of similar calculations for various $\Delta$ and $\kappa_2$, keeping $\kappa_1 = 0.06$ meV and $J=0.1$ meV constant, are presented in Fig.~\ref{fig5}. Figures~\ref{fig5}(a,b,c) correspond to a system with $\gamma=\kappa_1/3$, and Figs.~\ref{fig5}(d,e,f) correspond to $\gamma=0$. The dark red regions in Figs.~\ref{fig5}(a,d) indicate the parameters for which the reflectance vanishes.

Besides the numerical results, all panels of Fig.~\ref{fig5} display two analytical predictions for $\mathcal{R}_1=0$. The green dashed curve follows from Equation~\eqref{eq11V} with $\kappa_1 = 0.06$ meV and $J=0.1$ meV; this frequency-dependent solution, enabling $\mathcal{R}_1=0$ at any $\kappa_2$, is exclusive to coupled cavities. In contrast, the gray dashed line independent of $\Delta$ in all panels of Fig.~\ref{fig5} corresponds to a solution  where the two cavities effectively behave as a single one. In this case, the reflectance minimum takes place at $\kappa_2 = \kappa_1 - 2\gamma$, which can be recognized as the counterpart of Eq.~\eqref{eq:ZeroReflection_SKR} for a single cavity if one lets $\gamma \rightarrow 2\gamma$ considering that we have twice the intrinsic losses in the effective single cavity. Overall, our analytical and numerical results demonstrate that $\mathcal{R}=0$ can be achieved across a wide range of $\kappa_2$ by tuning $\Delta$. This tunability is impossible to achieve with a single cavity, where $\mathcal{R}=0$ only occurs for $\kappa_{1} =\Gamma/2$ regardless of $\Delta$.

We proceed to analyze the influence of $\gamma$ on $IR$ by comparing Figs.~\ref{fig5}(b) and~\ref{fig5}(e). Notice in Fig.~\ref{fig5}(e) that $IR\approx 1$ (black region in the color plot) at the values of $\kappa_2$ and $\Delta$ for which $\mathcal{R}_1=0$, along the green dashed line. Thus, for $\gamma=0$ there is negligible isolation when $\mathcal{R}=0$. In contrast, high $IR$ and $\mathcal{R}_1=0$ can be simultaneously achieved for a broad range of $\kappa_2$ when $\gamma \neq 0$. This is evidenced by the overlap of the green dashed line and the blue region of the colorplot in Fig.~\ref{fig5}(b). These results highlight how adding intrinsic loss $\gamma$ offers the possibility to tune the parameters ($\kappa_2$, $\Delta$) so as to achieve simultaneously suppressed reflectance and high non-reciprocity.

Simultaneously achieving high $IR$ and $\mathcal{R}_1=0$  by setting $\gamma \neq 0$ and tuning $\kappa_2$ requires $J>\kappa_1, \gamma$ \cite{NoteJ}. This result follows from Eq.~\eqref{eq11V}, which for $J<\kappa_1, \gamma$ leads to purely imaginary detunings $\Delta$. Physically, $J<\kappa_1,\gamma$ means that the two cavities act as if being decoupled. Consequently, the $\mathcal{R}_1=0$ solution that is only present in coupled cavities vanishes. In contrast, for  $J>\kappa_1,\gamma$ there is a finite range of $\kappa_2$ for which $\Delta$ is real; these are physically realizable  $\mathcal{R}_1=0$ states. Further increasing $J$  above $\kappa_1, \gamma$ enlarges the range of $\kappa_2$  and $\Delta$  over which high $IR$ and $\mathcal{R}_1=0$ can be simultaneously achieved. Note that while $J$ needs to be greater than both $\kappa_1$ and $\gamma$ for coupled cavity physics to emerge, $J$ can be much less than the total losses $2\gamma+ \kappa_1 + \kappa_2$ provided that Eq.~\eqref{eq11V} is satisfied.

The ability to simultaneously achieve high $IR$ and $\mathcal{R}_1=0$  by setting $\gamma \neq 0$ and tuning $\kappa_2$ comes at the expense of a degraded total transmission, as Figs.~\ref{fig5}(c,f) show. Whereas unitary transmission can be achieved for $\gamma=0$ [Fig.~\ref{fig5}(f)], for $\gamma= \kappa_1/3$ [Fig.~\ref{fig5}(c)] the transmittance is limited to a maximum value around 0.75. The trade-off between unitary transmission and  non-reciprocity for two identical coupled cavities also exists for a single cavity. However, by introducing intrinsic loss $\gamma$ in the coupled cavity system, high transmittance is traded for the ability to tune the parameters ($\kappa_2$, $\Delta$) leading to $\mathcal{R}_1=0$. In contrast, giving away high transmittance through a single cavity by introducing intrinsic loss does not enable one to tune any of the parameters to achieve $\mathcal{R}=0$.

\begin{figure*}[t]
	\includegraphics[width=\linewidth]{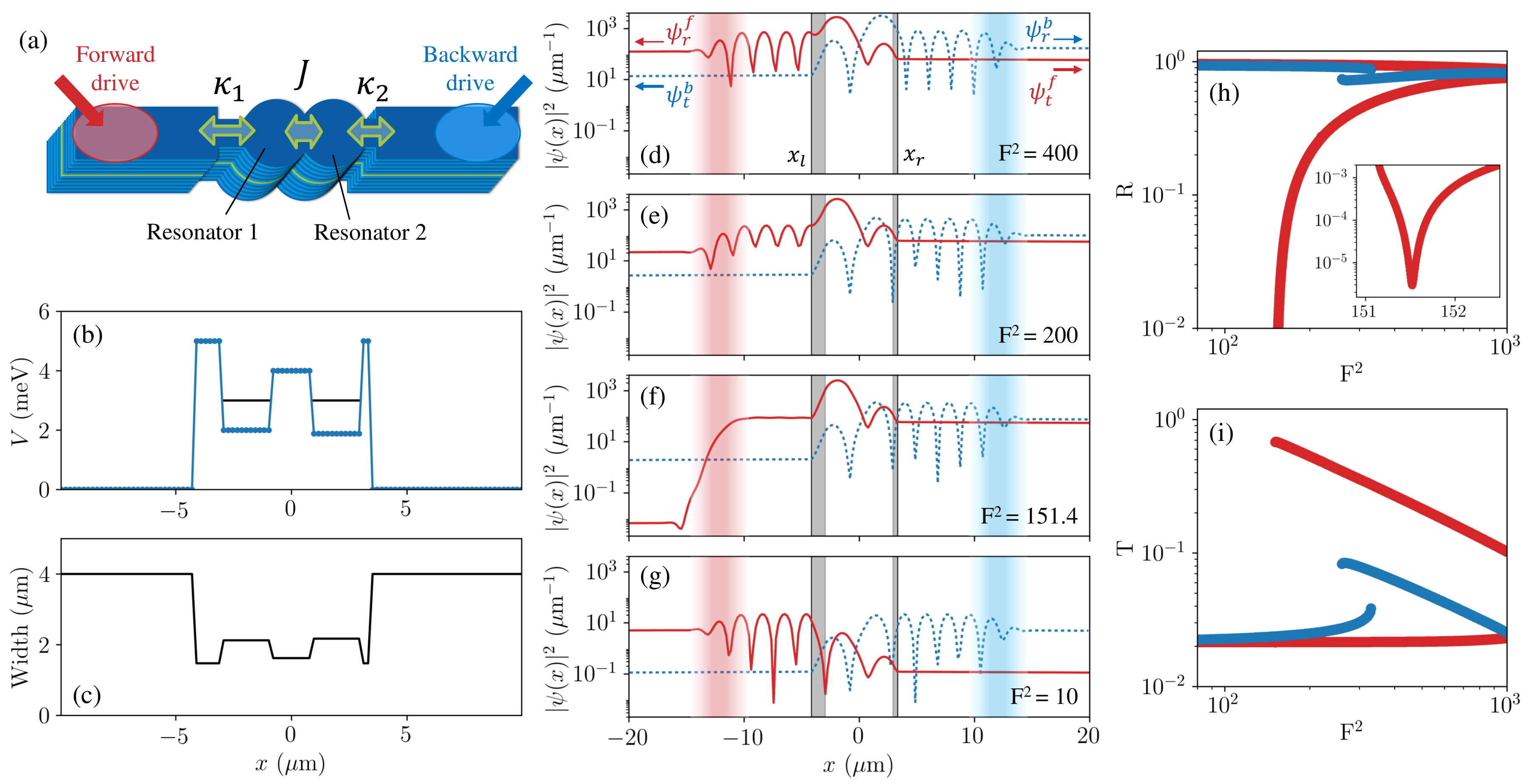}
	\caption{\label{fig6} (a) Sketch of the proposed implementation of coupled nonlinear cavities with asymmetric coupling to external ports based on etched polariton structures. (b) Potential $V(x)$ used in Eq.\eqref{eq:GPE}. Black lines indicate the energy $\hbar\omega_0$ of the confined modes in the wells. (c) Lateral width of the 2D structure realizing the potential energy landscape in (b). (d-g) Spatially resolved density $|\psi(x)|^2$, for $\omega = 3.49$ meV and a drive intensity indicated in the bottom right corner. Full red (dashed blue) line is for the forward (backward) configuration, with a drive on the left (right) of the wells. Shaded red and blue regions indicate the position of the drive in the forward and backward propagation direction, respectively. The positions of the external barriers are shown in gray, and their outer edge position is noted $x_{l,r}$ (solid black lines). (h, i) Reflectance and transmittance of the device, for $\omega = 3.49$ meV. The inset in (h) is a zoom on the reflectance dip, around the critical driving power $\mathrm{F_c^2}=151.4$}
\end{figure*}

\section{Driven-dissipative Gross-Pitaevskii calculations}

In this section we propose a design, based on polaritons in semiconductor microcavities, for the experimental implementation of non reciprocity and zero reflection using non-linearity in coupled resonators. Polaritons are quasi-particles arising from the strong coupling between excitons confined in a quantum well and photons in a cavity\cite{Weisbuch92}. Polaritons mutually interact due to their excitonic component, giving rise to strong Kerr-type optical nonlinearities~\cite{Carusotto13}. Several techniques are known for confining polaritons, and for coupling confined polariton modes~\cite{Abbarchi13, Dreismann14, Dufferwiel15, Urbonas, Deng15,  Rodriguez16, Deveaud17}.

We consider a microstructure schematically represented  in Fig. \ref{fig6}, which could be fabricated via deep etching of a planar microcavity. The microstructure is made of two coupled pillars linked in an asymmetric way to a one-dimensional waveguide via two constrictions. This 2D structure can be mapped to an effective 1D potential for polaritons. We make this mapping by considering that the lateral confinement creates a local potential inversely proportional to $w^2$, $w$ being the square of the structure width ~\cite{Dasbach2002, Nguyen2013}:
\begin{align}
V(x) = \hbar^2 / 2m (\pi / w(x))^2.
\label{eq:V2w}
\end{align}
For strong lateral confinement, the different transverse modes of the waveguide are far apart in energy. Thus, we can safely consider only the lowest energy band with an effective polariton mass $m$.

The evolution of the polariton wavefunction $\psi(x,t)$ in a potential landscape $V(x)$ is governed by the following driven-dissipative 1D Gross-Pitaevskii equation~\cite{Carusotto13, Gerace2012}:
\begin{align}
i \hbar \frac{\partial \psi (x,t)}{\partial t} =& \left(- \frac{\hbar^2}{2 m} \frac{\partial^2}{{\partial x}^2} + \hbar U |\psi (x)|^2 -i \frac{\hbar \gamma}{2} \right) \psi (x,t) \nonumber \\
& + V(x) \psi (x,t) + i \mathcal{F}(x) e^{-i (\omega t - k x)},
\label{eq:GPE}
\end{align}
where $\hbar U$ is the repulsive polariton-polariton interaction energy and $\gamma$ is the decay rate.
The last term in Eq.~\eqref{eq:GPE} corresponds to a monochromatic driving field of amplitude $\mathcal{F}$, frequency $\omega$, and wavevector $k$.
We compute the steady-state solutions of Eq.~\eqref{eq:GPE} with $m = 3 \times 10 ^{-5} \ m_{e}$ ($m_e$ is the free electron mass), $\hbar U = 0.3 \ \mathrm{\mu eV . \mu m}$ and $\hbar \gamma = 20$ $\mathrm{\mu eV}$. These values are taken from recent experiments~\cite{Rodriguez16, Goblot2016}.

Next, we explain how we tailor the potential $V(x)$ in order to realize the non-reciprocal coupled-cavity design. We target a coupling between left and right confined modes  $J = 100$ $\mu$eV,  and couplings to the waveguides $\kappa_1 = 60$ $\mu$eV and $\kappa_2 = 450$ $\mu$eV. According to Fig.~\ref{fig5}, these values should yield both zero reflectance and good isolation at the optimal $\Delta$ [given by Eq.~\eqref{eq11V}].

Our approach  to define $V(x)$ in relation to the zero-dimensional model is based on solving Eq.~\eqref{eq:GPE} for values of $F$ where the interaction term $\hbar U |\psi(x)|^2$ is negligible and the response is linear to a very good approximation. We begin by considering a potential landscape with a single well [corresponding to either of the two wells in Fig.~\ref{fig6}(b)]. For a well of length 2 $\mu$m, there is a single confined mode therein, with confinement energy $\omega_0 = 3.0$ meV. Next, we add a single potential barrier to tailor the coupling of the well to the waveguide. Leakage of polaritons from the well through the barrier broadens the confined mode linewidth. The coupling is extracted from this broadened linewidth. For a barrier height of 3 meV, $\kappa_1 = 60$ $\mu$eV is achieved with a barrier length $1.2$ $\mu$m, and $\kappa_2 = 450$ $\mu$eV is achieved with a length 0.4 $\mu$m. Finally, to design the height of the barrier between the two cavities to the desired value of the coupling $J$, we consider coupled wells similar to those in Fig.~\ref{fig6}(b) but with external barriers of effectively infinite thickness. The central barrier creates an effective coupling of amplitude $J$ for polaritons between the wells. This leads to bonding and antibonding modes with energies $\omega_0-J$ and $\omega_0+J$, respectively. We find that a barrier of height 2 meV and length 1.9 $\mu$m is required to get the desired $J = 100$ $\mu$eV. We assume that $J$ is not affected by the finite width of the barriers connecting the cavities to the 1D channels. The potential $V(x)$ resulting from the above design is shown in Fig.~\ref{fig6}(b), and the width of the corresponding 2D structure for an experimental implementation is shown in Fig.~\ref{fig6}(c). Note that in Fig.~\ref{fig6}(b) the reference for the potential ($V=0$) corresponds to the confinement potential in the waveguides, of width $4$ $\mu$m.

Forward and backward configurations are considered in order to determine the transmission and isolation properties of the device. The drive is either on the left of the double well and injects polaritons propagating towards the right (forward configuration), or the opposite (backward). These two situations are described in Eq.~\eqref{eq:GPE} with a drive term $\mathcal{F}_{f,b}(x) = F e^{-(x - x_{f,b})^2/2\sigma^2}$, corresponding to a spot of gaussian shape, centered on $x_{f, b} = \mp 12$ $\mu$m in the forward, backward configuration respectively. We choose a spot with 3 $\mu$m FWHM ($\sigma = 1.27$ $\mu$m). Additionally, the drive central wavevector $k_{f, b}$ is set to match the single polariton dispersion at energy $\hbar \omega$, to ensure efficient coupling with the modes in the external ports ($k_{f,b} = \pm \sqrt{2 m \omega}/\hbar$).

We now investigate the performance of the proposed device in terms of isolation and suppression of reflectance.
The steady-state density profiles in the wire, calculated for different drive intensities, are shown in Figs.~\ref{fig6}(d-g). The drive detuning $\Delta = \omega - \omega_0$ is fixed to $\Delta =$ 0.49 $\mathrm{meV}$ for all cases. In the forward configuration (full red line), we define the transmitted field $\psi_t^f(x)$ as the field $\psi(x)$ on the right side of the right external barrier ($x > x_r=3.35$ $\mu$m). In the backward configuration (dashed blue line), the transmitted field $\psi_t^b(x)$ is evaluated at the left of the left external barrier ($x<x_l=-4.15$ $\mu$m). The reflected field $\psi_r^{f,b}(x)$ is defined in a similar way. Two features in Fig.\ref{fig6}(d-g) characterize the reflected field intensity in the forward configuration: i) the polariton density $|\psi(x)|^2$ to the left of the pumping region ($x < -15\mathrm{\mu m}$), and ii) the density modulation between the pumping region and the left external barrier ($-10<x<-4.15\mathrm{\mu m}$), which results from the interference between the incident and scattered field.
Starting from a high drive intensity $F^2 = 400$ (Fig.~\ref{fig6}(d)), the difference of transmitted field intensity in the forward, backward configurations shows the non-reciprocal character of the device. However, the device is highly reflective, as indicated by the strong interference pattern for $-10<x<x_l$. Decreasing the drive intensity to $F^2 = 200$ (Fig.~\ref{fig6}(e)), we observe a reduced amplitude of the interference. Eventually, decreasing $F^2$ further, the interference has completely disappeared at $F^2 = 151.4$ (Fig.~\ref{fig6}(f)). This indicates a suppressed reflection, and correspondingly the polariton density is very low for $x<-15$ $\mu$m. Notice that non-reciprocal transmission is still observed at this critical drive $F_c^2 = 151.4$. This is no longer the case for a drive intensity $F^2 = 10$ (Fig.~\ref{fig6}(g)), i.e., in the linear regime.
In the absence of nonlinearities, the device has identical reflection and transmission properties whether in the forward or backward configuration.

To extract more quantitative information, we compute the transmittance and reflectance at a given $F$.  From the calculated intensity profiles we extract the transmitted, reflected, and incident fields intensities $|\psi_{t, r, i}|^2$ at the external barriers position $x_{l, r}$. For example, in the forward configuration $\psi_r^f(x_l)$ is computed by extrapolating to $x_l$ the slow exponential decay at the left side. The incident field is deduced from the interference pattern in the region $-10<x<x_l$ (similar procedure is used for the backward configuration).
Adapting the definition from the previous sections to the 1D model, the transmittance and reflectance are then given by $\mathcal{T},\mathcal{R} = |\psi_{t, r}|^2 / |\psi_{i}|^2$. Figures~\ref{fig6}(h,i) show the calculated $\mathcal{R},\mathcal{T}$ versus $F^2$ for $\Delta = 0.49$ meV, in the forward (red) and backward (blue) configuration. We obtain the features predicted by the 0D model (section II): a suppression of the reflectance down to less than $10^{-5}$ is observed when driving forward, at a critical drive $F_c^2=151.4$. Moreover, at this critical drive intensity, the forward transmittance is 0.68 while backwards  transmittance is 0.024. This corresponds to an isolation ratio of 28 at $F_c$.

% Figure 2
\begin{figure}[t]
	\includegraphics[width=\linewidth]{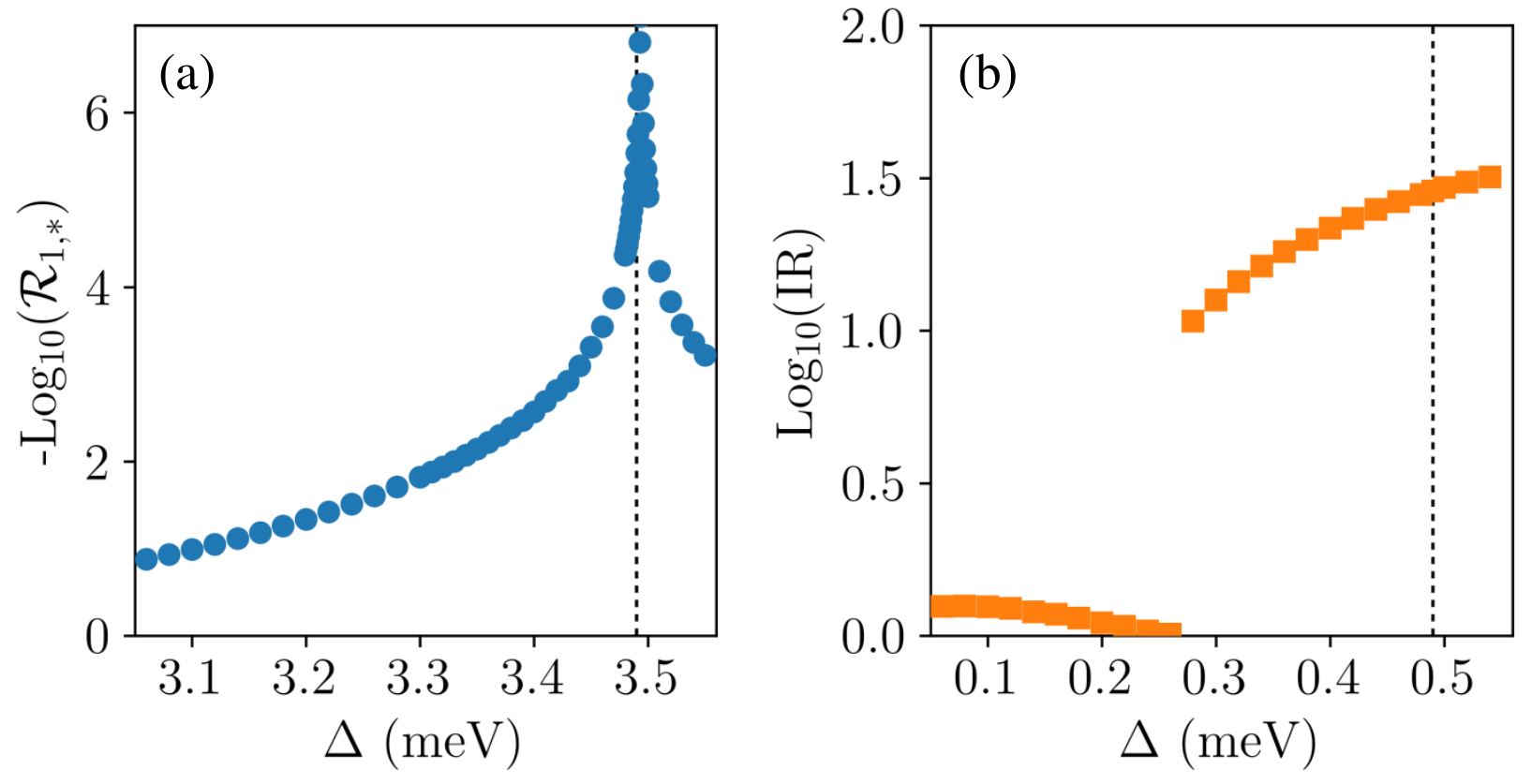}
	\caption{\label{fig7} (a) Minimum reflectance $\mathrm{\mathcal{R}_{1,*}}$ observed for any value of F, as a function of the detuning $\Delta$. (b) Isolation ratio calculated at the value of F for which $\mathrm{\mathcal{R}_{1,*}}$ was obtained. The dotted line indicates the detuning used in Fig. \ref{fig6}(d-h).}
\end{figure}

Similar to the discussion in the previous section, we extract the minimum value of $\mathcal{R}$ in the forward configuration when varying $F$, for different values of the drive energy detuning $\Delta$. We also compute the IR at $\mathcal{R}_{1,*}$. The results, presented in Fig. \ref{fig7}, show that $\mathcal{R}_{1,*}$ becomes arbitrarily small around $\Delta = 0.494$ meV for finer steps in $F$ and $\Delta$. Values of $\mathcal{R}_{1,*}$ below $10^{-7}$ were not reached in the 1D calculations due to numerical rounding errors.

We note that with the values of $\kappa_{1,2}$ extracted from our design, the analytical expression~\ref{eq11V} for the suppression of $\mathcal{R}$ in the 0D model gives $\Delta = 0.271$ meV. In the present case, we find $\mathcal{R}_{1,*}=0$ for $\Delta = 0.494$ meV. This difference  could be due to nonlinear spatial modifications of the modes in the double quantum wells. Indeed, we can see for instance in Fig.~\ref{fig6}(g) that the location of the density maxima within the two cavities depends on the excitation side. This distortion of the resonator modes, not captured by the 0D model, could modify the subtle interference effect responsible for $\mathcal{R}_{1,*}=0$. Nevertheless, the results from the 1D simulations show that our simple 0D model captures all of the key features discussed previously: suppression of $\mathcal{R}$ and high isolation ratio. Experimentally, since the value of $\kappa_2$ is often not tunable after fabrication, the presence of two coupled cavities ensures that $\mathcal{R}=0$ can be achieved by adjusting $\Delta$.

\section{Conclusion}
We have investigated the reflectance, transmittance, and non-reciprocity of single and coupled cavities with Kerr-type nonlinearity under continuous driving and dissipation. We derived analytical expressions predicting the existence of unidirectional zero-reflectance states, and we verified these predictions with numerical calculations based on nonlinear coupled mode theory and on the driven-dissipative Gross-Pitaevskii equation in one spatial dimension. We demonstrated how zero-reflection and high non-reciprocity can be simultaneously obtained by tailoring the leakage rates of the cavities to their input-output ports. For a single cavity, zero-reflection can only be achieved for one particular value of the input-output leakage rate difference. In contrast, for coupled cavities we have found that zero-reflection can be achieved for any value of the leakage rates provided that one can tune the operation frequency. Finally, we have presented the design of an experimental structure on which our predictions could be tested. A limitation of our approach to simultaneously obtain zero-reflection and non-reciprocity is that it is inherently limited in the operational power range, as expected due to the nonlinear origin of these effects.
Nevertheless, we expect these results to assist in the design of nonlinear optical isolators, and other devices where light is intended to propagate one-way only and with zero-reflection at the input port.

\section*{Acknowledgments}
This work is part of the research programme of the Netherlands Organisation for Scientific Research (NWO). S.R.K.R. acknowledges a NWO Veni grant. This work was supported by the ERC StG Honeypol, the EU-FET Proactive grant AQuS, the French National Research Agency (ANR) project Quantum Fluids of Light (ANR-16-CE30- 0021) and the Labex CEMPI (ANR-11-LABX-0007) and NanoSaclay (ICQOQS, Grant No. ANR-10-LABX- 0035), the CPER Photonics for Society P4S, and the M\'etropole Europ\'eenne de Lille via the project TFlight.

\section*{Appendix A: Complex fields of coupled cavities}
Here we explain how to solve equations~\eqref{cav1} and\eqref{cav2} to get the steady-state number of photons in the cavities $n_j$ and the complex fields $\psi_j$ ($j=1,2$). We start by rearranging equations~\eqref{cav1} and~\eqref{cav2} as follows:

\begin{align}
\psi_1 &=J^{-1}\left( -\Delta_2 - i \frac{\Gamma_2}{2} + U n_2 \right) \psi_2,  \label{al2cav1} \\
\psi_2 &=\left[ i \sqrt{\kappa_{1}} F + \left( -\Delta_1 - i \frac{\Gamma_1}{2} + U n_1 \right) \psi_1 \right] J^{-1}. \label{al2cav2}
\end{align}

To solve for $n_{1,2}$, we insert the expression for $\psi_1$ in Eq.~\eqref{al2cav1} into Eq.~\eqref{al2cav2}, and then multiply both sides with their complex conjugates. This leads to a polynomial equation (of order 9) in powers of $n_2$. Each root of that polynomial, subject to the physical condition $n_2>0$, corresponds to a steady-state. Next, we can use the solutions for $n_2$ and Eq.~\eqref{al2cav1} to calculate $n_1$.

We now seek expressions for the complex field $\psi_j$, which is related to $n_j$ via $\psi_j = \sqrt{n_j} e^{i\phi_j}$. Using this relation in Eqns.~\eqref{al2cav1} and~\eqref{al2cav2}, we arrive (after some algebra) to the following expressions for the phase factors,

\begin{align}
e^{i\phi_1} &= \frac{\sqrt{n_2}e^{i \phi_2}}{\sqrt{n_1} J} \left( -\Delta_2 - i \frac{\Gamma_2}{2} + U n_2 \right) , \label{ephi1} \\
e^{i\phi_2} &= \frac{J i \sqrt{\kappa_1} F}
{ \sqrt{n_2} \left[ J^2 - \left( -\Delta_1 - i \frac{\Gamma_1}{1} + U n_1 \right) \left( -\Delta_2 - i \frac{\Gamma_2}{2} + U n_2 \right)  \right]}, \label{ephi2}
\end{align}

\noindent from which the complex fields $\psi_j$ can be constructed once the $n_j$'s are known.

\section*{Appendix B: Analytical expression for zero-reflection in coupled cavities}
In this section we derive an analytical expression guaranteeing the existence of a zero-reflection state in coupled cavities. We begin the derivation by inserting the expression for $\psi_2$ in Eq.~\eqref{cav2} into the expression for $\psi_1$ in Eq.~\eqref{cav1}. After rearranging, we get:

\begin{equation} \label{eq4V}
0 = \left( - \Delta + U n_1 - i \frac{\Gamma_1}{2} \right) \psi_1 - \frac{J^2 \psi_1}{- \Delta + U n_2 - i \frac{\Gamma_2}{2}} + i \sqrt{\kappa_1} F
\end{equation}

According to Eq.~\eqref{eq:R2cavs}, $\mathcal{R}=0$ implies $F= \sqrt{\kappa_1} \psi_1$. Hence, let us insert this expression for $F$ into ~\eqref{eq4V}, and separate the real and imaginary parts. The equation for the real parts reads,

\begin{equation} \label{eq5V}
0 =  -\Delta + Un_1 - \frac{J^2}{ (- \Delta + U n_2)^2 + \frac{\Gamma_2^2}{4}} (- \Delta + U n_2).
\end{equation}

%\begin{equation} \label{eq6V}
%0 = \frac{-\Gamma_1}{2} - \frac{J^2 \Gamma_2 /2}{ (- \Delta + U n_2)^2 + \frac{\Gamma_2^2}{4}} + \kappa_1.
%\end{equation}
%Defining $\tilde{\Gamma_1} = \kappa_1 - \gamma$, Eq.~\eqref{eq6V} becomes

\noindent Meanwhile, the equation for the imaginary parts reads,
\begin{equation} \label{eq7V}
0 = \tilde{\Gamma_1} - \frac{J^2}{ (- \Delta + U n_2)^2 + \frac{\Gamma_2^2}{4}} \Gamma_2,
\end{equation}

\noindent where we have defined $\tilde{\Gamma_1} = \kappa_1 - \gamma$.

Let us now rewrite Eq.~\eqref{cav2}, in order get the following relation between the number of photons in the cavities:

\begin{equation} \label{eq3V}
\frac{n_2}{n_1} = \frac{J^2}{ ( - \Delta + U n_2)^2 + \frac{\Gamma_2^2}{4} }.
\end{equation}

\noindent On the one hand, inserting Eq.~\eqref{eq3V} into Eq.~\eqref{eq5V} leads to

\begin{equation} \label{eq8V}
( - \Delta + U n_1)n_1 = ( - \Delta + U n_2)n_2.
\end{equation}

\noindent On the other hand, combining Eq.~\eqref{eq3V} and Eq.~\eqref{eq7V} yields

\begin{equation} \label{eq9V}
\frac{n_2}{n_1} = \frac{\tilde{\Gamma_1}}{\Gamma_2}.
\end{equation}

Using Eq.~\eqref{eq8V} and Eq.~\eqref{eq9V}, we can now solve for $n_1$ and $n_2$. We use Eq.~\eqref{eq9V} to substitute $n_1$ in Eq.~\eqref{eq8V} and get the following equation for $n_2$:

\begin{equation} \label{eq10bV}
0 = \left(1 - \frac{\Gamma_2}{\tilde{\Gamma_1}} \right) \left( - \Delta + U \left(1 + \frac{\Gamma_2}{\tilde{\Gamma_1}} \right) n_2 \right) .
\end{equation}

One solution for Eq.~\eqref{eq10bV} is $\Gamma_2 = \tilde{\Gamma_1}$, which is equivalent to $\kappa_1 = \kappa_2 + 2\gamma$. As discussed in the main text, this corresponds to the solution for $\mathcal{R}=0$ of a single resonator with intrinsic loss $2\gamma$, as given by Eq.~\eqref{eq:ZeroReflection_SKR}. Note that in this case, Eq.~\eqref{eq9V} imposes $n_1=n_2$, i.e. equal population is both resonators, confirming that they behave as a single one.

Coming back to Eq.~\eqref{eq10bV}, for $\Gamma_2 \neq \tilde{\Gamma_1}$ we get the expression for $n_2$ corresponding to a second branch of solution

\begin{equation} \label{eq10V}
n_2 = \frac{\Delta}{U} \frac{\tilde{\Gamma_1}}{\tilde{\Gamma_1} + \Gamma_2} .
\end{equation}

Finally, we insert Eq.~\eqref{eq10V} into Eq.~\eqref{eq7V}, and after a little bit of algebra we get

\begin{equation} \label{eq11Vb}
\Delta^2 = (\kappa_1 + \kappa_2)^2 \left( \frac{J^2}{(\kappa_1 - \gamma)(\kappa_2 + \gamma) } - \frac{1}{4} \right) .
\end{equation}

Note that we can also determine the value of the drive amplitude $F_c$ corresponding to the above solution. To this end, we recall that $\mathcal{R}_{1}=0$ imposes $F = \sqrt{\kappa_1} \psi_1$. $F$ is a real number, so in this case $\psi_1$ is also real and we have $\psi_1 = \sqrt{n_1}$. We use Eq.~\eqref{eq9V} and Eq.~\eqref{eq10V} to obtain the expression for $n_1$ and finally get the following expression for $F_c$:
\begin{equation} \label{eq11Vf}
F_c^2 = \frac{\Delta}{U} \frac{\kappa_1 (\kappa_2 + \gamma)}{\kappa_1 + \kappa_2}.
\end{equation}

Equation~\eqref{eq11Vb} constraints the parameters $\Delta$, $J$, and $\kappa_{1,2}$ such that
$\mathcal{R}_{1}=0$. Equation~\eqref{eq11Vf} gives the drive $F_c$ at which $\mathcal{R}_{1}=0$ is achieved.

In particular, when $\gamma=0,$ Eq.~\eqref{eq11Vf} is symmetric to the switching of input ports $1 \leftrightarrow 2$. Exciting from each side leads to $\mathcal{R}=0$, $\mathcal{T}=1$: in this particular situation the device is perfectly reciprocal.


\begin{thebibliography}{46}
\bibitem [{Not()}]{NoteJ}%
  \BibitemOpen
  \href@noop {} {}\bibinfo {note} {Strictly speaking, the lower bound on J
  given by Eq.~\eqref{eq11V} is $J^2 > (\kappa_1 - \gamma) \gamma / 4$, which
  is satisfied when $J > \kappa_1, \gamma$.}\BibitemShut {Stop}%
\end{thebibliography}
\end{document}